\documentclass[journal]{IEEEtran}
\usepackage{ifpdf}

\usepackage{cite}

\ifCLASSINFOpdf
\usepackage[pdftex]{graphicx}
\graphicspath{{Graphics}}
\else
\usepackage[dvips]{graphicx}
\graphicspath{{Graphics}}
\fi
\usepackage[intlimits]{amsmath}
\usepackage{esint}
\usepackage{verbatim}
\usepackage{color}

\usepackage{array}
 \usepackage{dblfloatfix}

\ifCLASSOPTIONcaptionsoff
  \usepackage[nomarkers]{endfloat}
 \let\MYoriglatexcaption\caption
 \renewcommand{\caption}[2][\relax]{\MYoriglatexcaption[#2]{#2}}
\fi
\usepackage{url}

\usepackage{bm}
\usepackage{color}

\usepackage{empheq} 
\usepackage{subfigure}

\newcommand{\e}{\mbox{e}}

\renewcommand{\vec}{\bm}
\newcommand{\be}{\begin{equation}}
\newcommand{\ee}{\end{equation}}
\newcommand{\bea}{\begin{eqnarray}}
\newcommand{\eea}{\end{eqnarray}}

\newcommand{\jb}{\mathrm{j}}

\begin{document}
%
\title{High Speed High Signal-to-Noise Ratio Antenna Measurements  --- Demonstration for UAV-Based Near-Field Measurements of Modulated Terrestrial Navigation Signals}
	
%
%
%
\author{Thomas~F.~Eibert, Denis Unruh, Thomas Mittereder, and Alexander~H.~Paulus

\thanks{The authors are with the Department of Electrical Engineering, School of Computation, Information and Technology, Technical University of Munich, 80290 Munich, Germany (e-mail: eibert@tum.de).
The work was supported by the German Federal Ministry for Economic Affairs and Energy under Grant 20E2124A. 
}}

\maketitle

\begin{abstract}
Antenna measurements with high signal-to-noise ratio (SNR) require long measurement or integration times of the receiver and can, thus, lead to a very long duration of the measurements, especially if many frequency and spatial samples need to be collected. In order to speed up such measurements, an approach is presented, which collects all measurement samples with short measurement times, performs a Fourier transform of the measurement samples, bandpass filters the desired measurement signals with a small bandwidth, and obtains high-SNR measurement samples according to the short measurement times by inverse Fourier transform. This approach can be utilized with single-frequency continuous wave (CW) transmit signals, but also with transmit signals containing several discrete frequency components, as, e.g., found for periodically modulated CW carriers. The approach is first worked out and demonstrated for simulated test data. Next, it is utilized for the processing of modulated near-field (NF) measurement data collected via an uninhabited aerial vehicle (UAV) at a Doppler high-frequency omnidirectional radio range (DVOR) and at the localizer of an instrument landing system (ILS). The extracted CW NF data is transformed into the far field (FF) and diagnostic information is obtained from the underlying inverse source solutions. 
\end{abstract}


%
\IEEEpeerreviewmaketitle

\section{Introduction}
\IEEEPARstart{A}{ntenna} measurements can be very time-consuming, especially if many frequency samples are needed and even more if very many spatial measurement samples have to be collected, as in the case of near-field (NF) antenna measurements \cite{Parini2020,IEEE2021,IEEE2012}. An important limiting factor for measurement speed is of course the finite speed of mechanical positioning systems, in particular in stepped measurement modes, where all frequency samples are collected for stationary antenna under test (AUT) and probes. However, even with continuously moving positioner or switched probe arrays, the measurement speed is limited by the measurement time of the radio frequency (RF) measurement system. In order to obtain a measurement sample at a given frequency with a certain signal-to-noise ratio (SNR), a certain measurement time is needed. This time corresponds to the inverse of the measurement bandwidth, which is the intermediate frequency (IF) bandwidth in a heterodyne receiver, such as used in a vector network analyzer (VNA). In a time-domain receiver, this measurement time becomes obvious as the integration time needed for the corresponding short-time Fourier transform into the frequency domain, which is still the standard domain for antenna measurements and corresponding field transformations~\cite{Parini2020,IEEE2021,IEEE2012,Yaghjian1986}. In order to perform antenna measurements with high SNR, the IF or measurement bandwidth must be chosen small, leading to a long measurement time per frequency sample. In a stepped measurement mode, the waiting time in every measurement position must be so large that all frequency samples can be collected, and with a continuously moving positioner system, the speed of the movement must be small enough such that the measurement signal can be assumed stationary within the measurement time of one frequency sample. 

Formally, such measurements can be accelerated by working with transmit (Tx) signals containing a collection of discrete frequencies, as, e.g., also found for certain periodically time-modulated signals. Transforming the related signals into the frequency domain allows again to treat the different continuous wave (CW) signals separately, e.g., by a corresponding field transformation. The advantage of such an approach is that the measurement or integration time is now only needed once for all frequencies contained in the signal, but it must here at least be so long that the individual frequency contributions can be resolved and extracted. Remarkably, such measurements are not really around in antenna measurements, which might be due to the fact that the commonly utilized VNAs only support CW Tx signals. Moreover, multi-frequency signals can have strong temporal magnitude variations, which are not that easy to handle by the measurement system. 
Similar is the situation for time-domain measurements with very short impulses, which have the potential to achieve very fast measurement speeds for wideband signals even with good SNR \cite{Jongh1997,Serhir2012}. Such measurements received quite some attention many years ago. However, the generation and the handling of the short impulses with large enough signal magnitudes and corresponding linearity problems is in general so problematic that impulse measurements are typically restricted to those cases where they are really mandatory~\cite{Hassett2011}.  

The concept of measurements with multiple-frequency signals is of course also relevant, if the measurements need to work directly with the modulated operational signals of communication, sensing, or navigation systems. In \cite{Faul2019,Faul2023}, NF measurements of continuously modulated signals of terrestrial navigation systems have been considered, where in addition to the already described so-called long-time measurement approach, also a so-called short-time measurement approach was introduced and demonstrated. In this short-time measurement approach, it is assumed that the modulation signal frequencies are considerably smaller than the carrier frequency and the measurements are performed with a VNA working with an IF bandwidth much larger than the modulation signal frequencies. As such, the modulation signal can be assumed constant within the rather short measurement time and the modulation signal is kind of sampled by the performed short-time measurements. With a periodic modulation signal, the measurement samples can be sorted according to their modulation states and CW field transformations can be performed for sub-sets of the measurement samples belonging to identical modulation states. Such an approach works well for continuously modulated signals with small modulation frequencies, however, the achievable SNR is here limited by the short measurement time, which must be so short that the modulation signal can be assumed constant. 

Considering signals with very different time scales, i.e., a fast time and a slow time, is actually very common in the characterization of mobile communication signals or channels~\cite{Bello1963,Parsons2000}. When considering modulated signals with small modulation frequencies, the carrier depends on fast time related to its high frequency and the modulation signal depends on slow time related to its considerably smaller frequencies. In such a situation, Fourier transforms can be performed separately with respect to fast time and slow time. In our measurement problem, performing the measurements with a short measurement time corresponding to a large measurement bandwidth around the carrier frequency, corresponds to performing a Fourier transform with respect to fast time, whose result is still dependent on slow time according to the modulation signal. Performing next a Fourier transform with respect to the slow time, i.e., with a much longer integration time than before, we obtain now the frequency dependent spectrum of the modulation signal and we can detect and extract the contained frequency components in a similar way as in the long-time measurement approach as mentioned before. As such, this two-step approach does not have much benefit compared to the long-time approach. However, if we separate the individual frequency components contained in this short-time frequency spectrum by bandpass (BP) filtering and perform separate inverse Fourier transforms for each of them, then we obtain CW frequency samples for all of them with short-time step size, but with the high SNR according to the narrow bandwidth of the BP filter. Based on this concept, high speed high SNR antenna measurements can be realized, where CW signals or modulated signals containing several discrete frequency components can be considered. If the utilization of Tx signals containing several discrete frequency components is not desired, the concept can also be used to speed up conventional CW measurements, where the Tx signal contains just one frequency component.  

Filtering techniques related to the method proposed here---as well as even more advanced methods---are, of course, well known in various fields, where measured signals are processed, see, e.g.,~\cite{Wiener1949,Kailath1974,Widrow1975,Hu2021}. In the field of antenna measurements, however, it is obviously all too tempting to perform the measurements directly with a narrow measurement bandwidth—--even though this results in long measurement times.    

In Sections~\ref{model} and \ref{simu} of this article, the high speed high SNR signal model and measurement concept are worked out and demonstrated by simulations.  
In Section~\ref{results}, the measurement and signal processing approach will be demonstrated for NF measurements of modulated terrestrial navigation signals, which have been collected via an uninhabited aerial vehicle (UAV). Some conclusions are drawn in Section~\ref{conclusions}.

\section{Measurement Signal Model}
\label{model}
\begin{figure}[t]
	\centering
	\subfigure[~]{\includegraphics[scale=0.47,keepaspectratio]{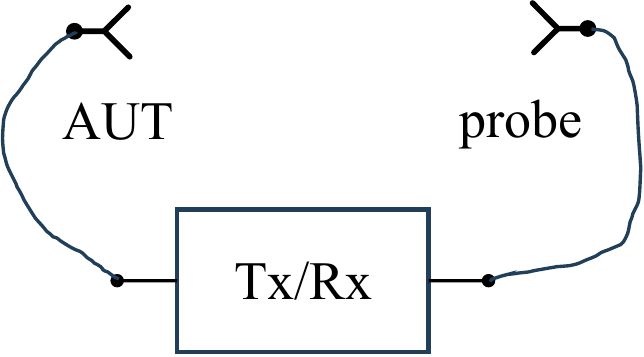}}\\[1.9mm]
	\subfigure[~]{\includegraphics[scale=0.47,keepaspectratio]{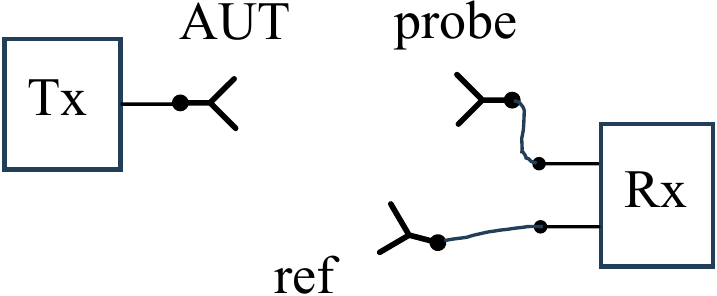}}\\
	\caption{Considered antenna measurement configurations. (a)~Measurement with coherent Tx/Rx unit. (b)~Coherent measurement with reference antenna (ref).}
	\label{conf}
\end{figure}

Let us consider coherent antenna measurement setups as illustrated in Fig.\,\ref{conf}.
The configuration in Fig.\,\ref{conf}(a) is the standard setup with a coherent transmit/receive (Tx/Rx) unit such as a VNA, where the Tx signal is fed into the AUT and the Rx signal is received by one (or possibly also more probe antennas). Due to reciprocity, the Tx signal could also be fed into the probe and the Rx signal would then be received by the AUT. In configuration Fig.\,\ref{conf}(b), the AUT transmits a signal, which is not phase-locked to the Rx, but coherent measurements can still be performed by measuring the probe signal coherently together with a reference signal received by a reference antenna (ref), which is stationary with respect to the AUT. Again, due to reciprocity, the roles of Tx and Rx could be flipped. 

For the description of the signal model, we work with complex base band signals $x(t)$ and $y(t)$, where $x(t)$ represents the Tx signal for the configuration in Fig.\,\ref{conf}(a) or the reference signal in Fig.\,\ref{conf}(b), and $y(t)$ represents the probe signal in both cases. The complex base band signals depend only on the slow time $t$. The connection to the corresponding radio frequency (RF) signals is given by in-phase quadrature (IQ) mixing with a CW carrier or local oscillator (LO) signal with frequency $\omega_{\mathrm{ LO}}$ representing the fast time dependence. In a practical implementation with a VNA, the RF signals are typically detected at an appropriately chosen IF frequency. This is advantageous with respect to the noise behavior and measurement accuracy, but delivers formally also the complex base band signals dependent on its slow time, if the IF bandwidth is appropriately chosen. 

Let us now consider measurements with a signal $x(t)$ consisting of $K$ possibly modulated CW signals given as
\be
x(t) = \sum_{k=1}^K x_k(t) = \sum_{k=1}^K m_k(t) \,\e^{\jb\omega_k t}\,\label{xsig}
\ee 
with the CW angular frequencies $\omega_k = 2\pi f_k$ and the also complex modulation signal $m_k(t)$, which can just be a complex constant for signals without modulation. The angular frequencies $\omega_k$ represent the different simultaneously measured RF frequencies $\omega_\mathrm{LO}+\omega_k$ all lying within the measurement or IF bandwidth of the receiver.
With this, the probe signal $y(t)$ becomes
\be
y(t) = \sum_{k=1}^K y_k(t) = \sum_{k=1}^K h_k(t)\, m_k(t) \,\e^{\jb\omega_k t}\,,
\ee
where the channel transfer functions
\be
h_k(t) = y_k(t)/x_k(t)
\ee
for the different CW frequencies are the actual measurement quantities of interest. At this point, it should be noted that the assumed separable, periodic, and smooth modulation signals $m_k(t)$, which cancel out, can, of course, only represent certain types of modulation such as amplitude or frequency modulation (AM or FM). 

In order to extract the $h_k(t)$ out of $y(t)$ and $x(t)$, we may perform a Fourier transform of both signals, where the goal is, however, to maintain the time dependence of $h_k(t)$. In a conventional approach, one would apply an appropriate sliding temporal window to both signals first, where the window length $T_{\mathrm W}$ is short enough to assume the $h_k(t)$ close to constant within the window length, but also long enough to accurately resolve the discrete frequencies due to the periodic modulation signals. With a window centered around a time sample $t_0$, we obtain 
\begin{eqnarray}
	X(t_0,\omega) &=& \sum_{k=1}^K \sum_{l=1}^{L_k} M_{kl}(t_0) \,\delta(\omega - \omega_{kl})\,,\\
	Y(t_0,\omega) &=& \sum_{k=1}^K \sum_{l=1}^{L_k} h_{kl}(t_0) M_{kl}(t_0) \,\delta(\omega - \omega_{kl})\,,
\end{eqnarray}
where $X(\omega)={\cal FT} \{x(t)\}$ is the Fourier transform of $x(t)$ and $\delta(.)$ is the Dirac delta distribution. Obviously, each of the $K$ CW signals contained in the measurement signal separates into $L_k$ discrete frequencies due to its periodic modulation, and the desired measurement quantities $h_{k,l}(t_0)$ can be easily extracted as the corresponding pre-factors. However, the key disadvantage of this long-time measurement approach is the long windowing or measurement time $T_{\mathrm W}$, which is needed to separately extract the desired measurement quantities for all the relevant frequencies. In particular, the channel transfer functions $h_{k,l}(t)$ are not allowed to change noticeably during $T_{\mathrm W}$, which can be a major limitation especially in continuously moving positioner systems. 

In order to overcome this problem, our approach is to perform the Fourier transform for the entire duration of the signals, or at least for very long subsections, resulting into
\begin{eqnarray}
X(\omega) &=& \sum_{k=1}^K \sum_{l=1}^{L_k} M_{kl}\,\delta(\omega - \omega_{kl})\,,\label{xf}\\
Y(\omega) &=& \sum_{k=1}^K \sum_{l=1}^{L_k} M_{kl}\, H_k(\omega)\, ** \,\delta(\omega - \omega_{kl})\,,\label{yf}
\end{eqnarray}
where $**$ means convolution and where the time variation of the channel transfer functions $h_k(t)$ is now appropriately considered via its Fourier transforms $H_k(\omega)$. Obviously, the bandwidth of the $H_k(\omega)$ must be so small that their convolution with the discrete carriers does not produce any overlaps in the overall spectral representation of $Y(\omega)$. If this is guaranteed, the spectra around the individual carriers can be extracted by multiplying with a spectral BP filter function $BP(\omega-\omega_{kl})$ and the desired measurement quantities dependent on the (slow) time $t$ related to the individual discrete frequencies contained in the measurement signal can be obtained by inverse Fourier transform giving
\begin{eqnarray}
	x_{kl}(t) &=& {\cal FT}^{-1}\{M_{kl}\,\delta(\omega - \omega_{kl})\}\,,\notag\\
	 &=& M_{kl}\, \e^{\jb\omega_{kl} t}\,,\label{xt}\\
	y_{kl}(t) &=& {\cal FT}^{-1}\{ M_{kl}\, H_k(\omega)\, ** \,\delta(\omega - \omega_{kl})\}\,,\notag\\
		 &=& M_{kl}\, h_{kl}(t)\,\e^{\jb\omega_{kl} t}\,\label{yt}
\end{eqnarray}
and finally taking
\be
h_{kl}^{\mathrm{BP}}(t) =  y_{kl}(t)/ x_{kl}(t)\,,\label{hsig}
\ee
where it was assumed that the BP filter function does not modify the desired signal.
With this, the time dependencies of the channel transfer functions related to all relevant discrete frequencies contained in the measurement signal are correctly recovered even though the relevant measurement or IF bandwidth as defined by the BP filter function $B(\omega-\omega_{kl})$ is very narrow and would require a long measurement time without allowable change of $h_{kl}(t)$ in a conventional approach. In the proposed approach with its additional inverse Fourier transform, the actual measurement times are chosen very short, corresponding to a large measurement or IF bandwidth covering at least all signal contributions contained in the signal $x(t)$ in (\ref{xsig}), and the very narrow bandwidths of the final measurement signals in (\ref{hsig}), which are relevant for the achievable SNRs, are completely obtained by the described signal processing procedure.

\section{Numerical Implementation and Signal Simulations}
\label{simu}
In a numerical implementation, all signals are available in a time discrete form, where we assume equidistant sampling with a temporal step size $\Delta t$. With this, discrete Fourier transforms and its inverse can be efficiently computed via fast Fourier transforms (FFTs). Whenever a Fourier transform is applied to a sequence of time samples, as, e.g., in (\ref{xf}) and (\ref{yf}), a Blackman window function~\cite{Harris1978} is first multiplied to the complete sequence in order to remove corresponding truncation effects. After a spectral representation of a signal is transformed back into the time domain, as, e.g., in (\ref{xt}) and (\ref{yt}), the complete sequence is divided by the earlier applied window function and a certain number of temporal samples is dropped at the beginning and end of the obtained sequence, again to remove the corresponding truncation effects. For the spectral filter function $B(\omega-\omega_{kl})$, we use a constant section in the middle of the filter combined with half flat-top windows at the band edges. In order to estimate the SNR of the relevant signals, we use a sliding averaging process with averaging time $T_{\rm av}$ according to
\be
SNR\{x(t)\} = \frac{x^2_0(t)}{\underset{T_{\rm av}}{\mathrm{mean}}\{|x(t) - x_0(t)|^2\}}
\ee
with 
\be
x_0(t) = \underset{T_{\rm av}}{\mathrm{mean}}\{x(t))\}\,,
\ee
which is implemented for the time discrete signal representations by considering a carefully determined number of discrete samples around a certain time sample.

\begin{figure}[t]
	\centering
	\subfigure[~]{\includegraphics[scale=0.42,keepaspectratio]{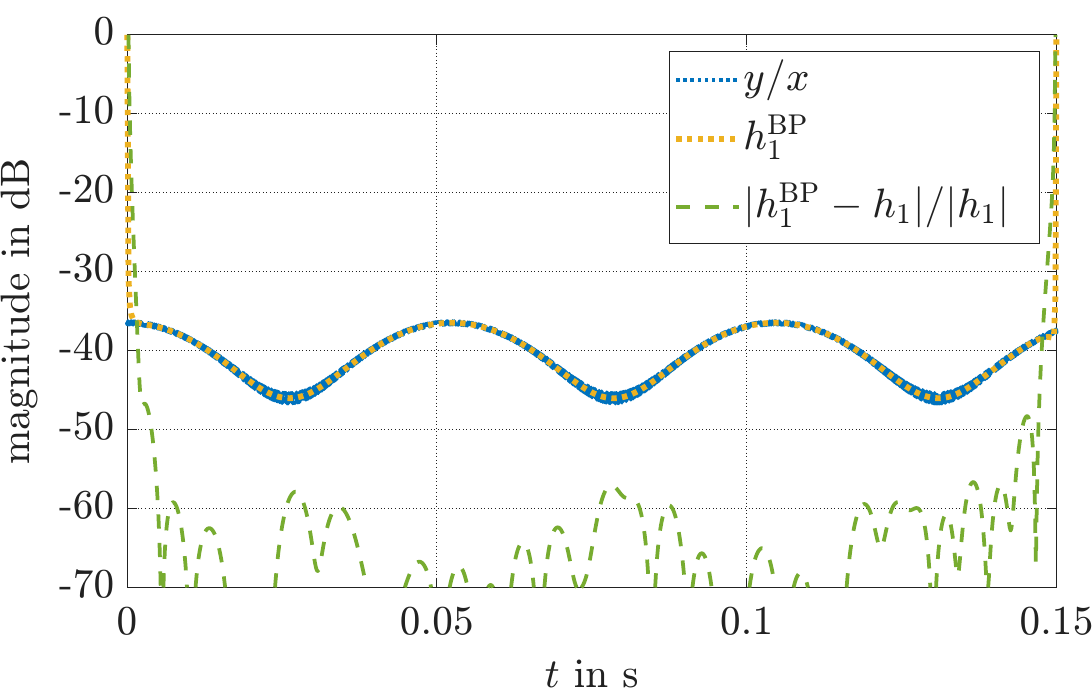}}\\[1.9mm]
	\subfigure[~]{\includegraphics[scale=0.42,keepaspectratio]{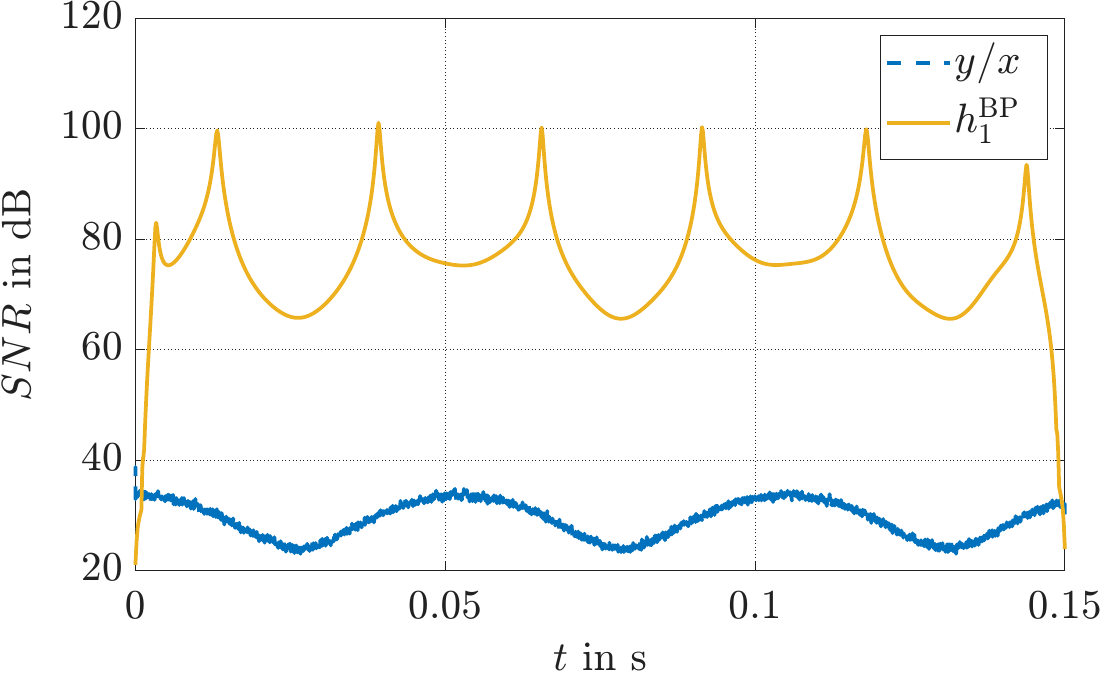}}\\
	\caption{Simulated measurement data with one CW signal. (a)~Noisy measurement signal and extracted channel transfer function together with its deviation from the reference. (b)~SNR of raw measurement data and of the BP-filtered transfer function.}
	\label{sim1}
\end{figure}

In a first simulation example, we use just one CW signal with a base band frequency $f_1 = 0$\,Hz and $m_1(t)=1$, i.e., with an RF frequency identical to the mixing carrier frequency. This signal is sampled in the base band with a time step $\Delta t = 1$\,$\mu$s corresponding to an IF or measurement bandwidth of 1\,MHz. Both $x(t)$ and $y(t)$ are superimposed with random noise and the channel transfer function is assumed to be
\be
h_1(t)= 0.005\, (2 + \cos(120\, t/\mathrm{s}))\,.
\ee
The relevant simulation results are depicted in Fig.\,\ref{sim1}. The raw measurement signal corresponding to the measurement bandwidth of 1\,MHz is $y(t)/x(t)$ and the BP-filtered measurement signal obtained by the presented signal processing methodology with a bandwidth of 300\,Hz is $h_1^{\mathrm{BP}}(t)$. Both are found in Fig.\,\ref{sim1}(a), where the observed relative deviation of $h_1^{\mathrm{BP}}(t)$ with respect to the reference $h_1(t)$ on the order of about $-60$\,dB and below corresponds approximately to the worst-case SNR of $h_1^{\mathrm{BP}}(t)$ as found in Fig.\,\ref{sim1}(b). The worst-case SNR improvement due to the BP filtering as can be observed in Fig.\,\ref{sim1}(b) corresponds pretty well with the expected value of  $10\log_{10}(1\,\mathrm{MHz}/300\,\mathrm{Hz}) = 35.2$\,dB. The strong peaks of the SNR in the slope regions of $h_1^{\mathrm{BP}}(t)$ are just artifacts in the SNR estimation.

\begin{figure}[t]
	\centering
	\subfigure[~]{\includegraphics[scale=0.43,keepaspectratio]{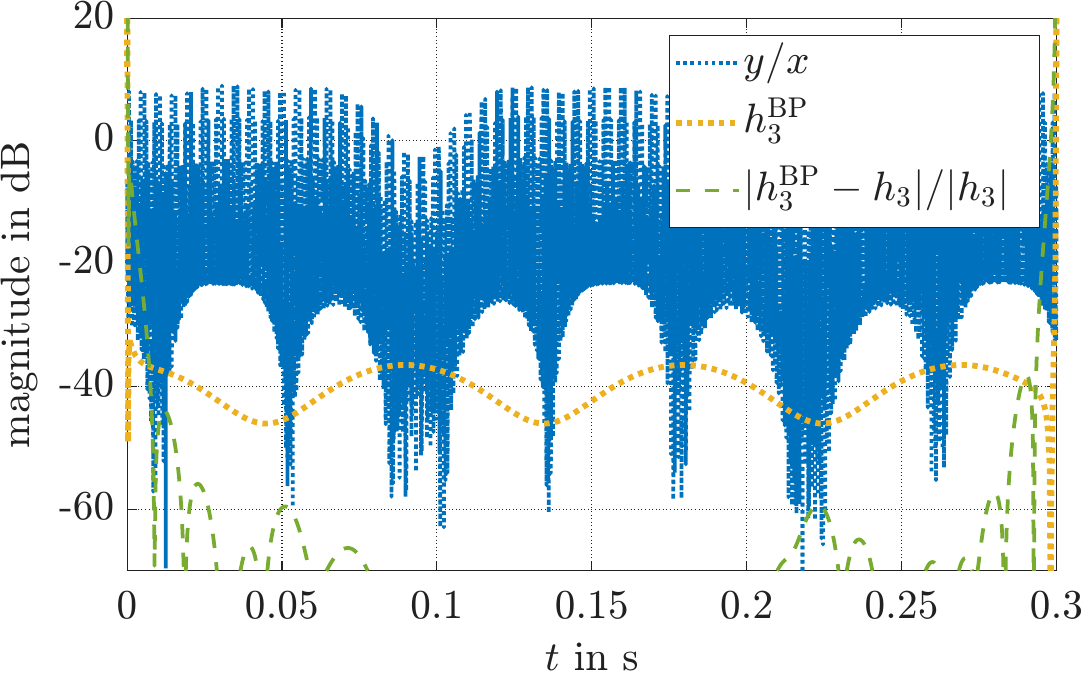}}\\[1.9mm]
	\subfigure[~]{\includegraphics[scale=0.43,keepaspectratio]{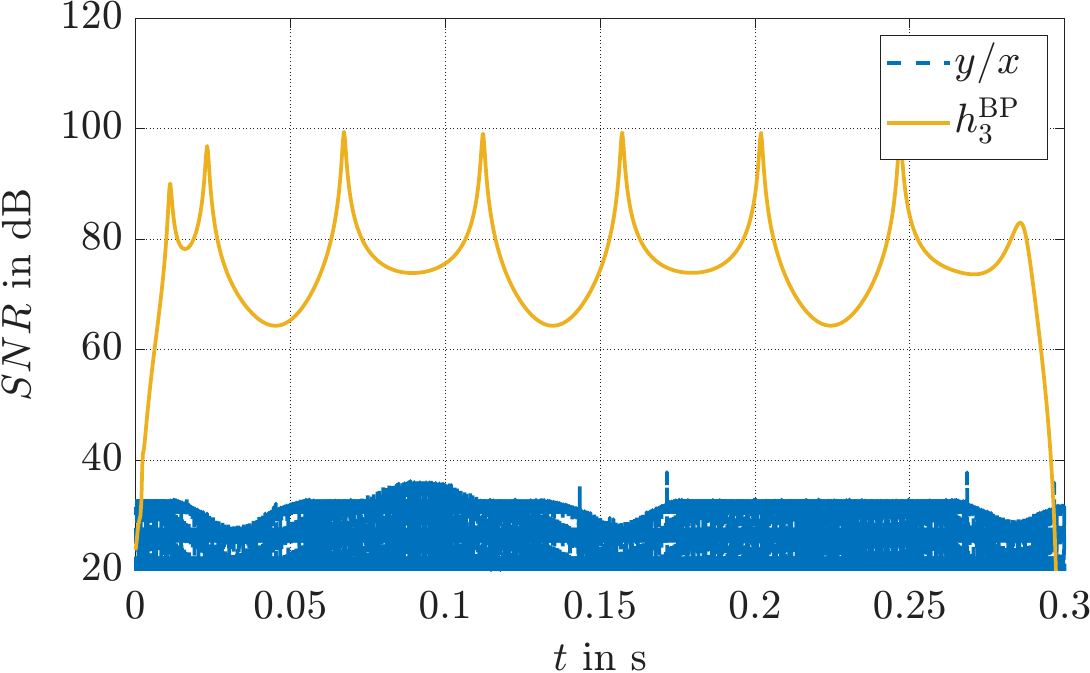}}\\[1.9mm]
	\subfigure[~]{\includegraphics[scale=0.43,keepaspectratio]{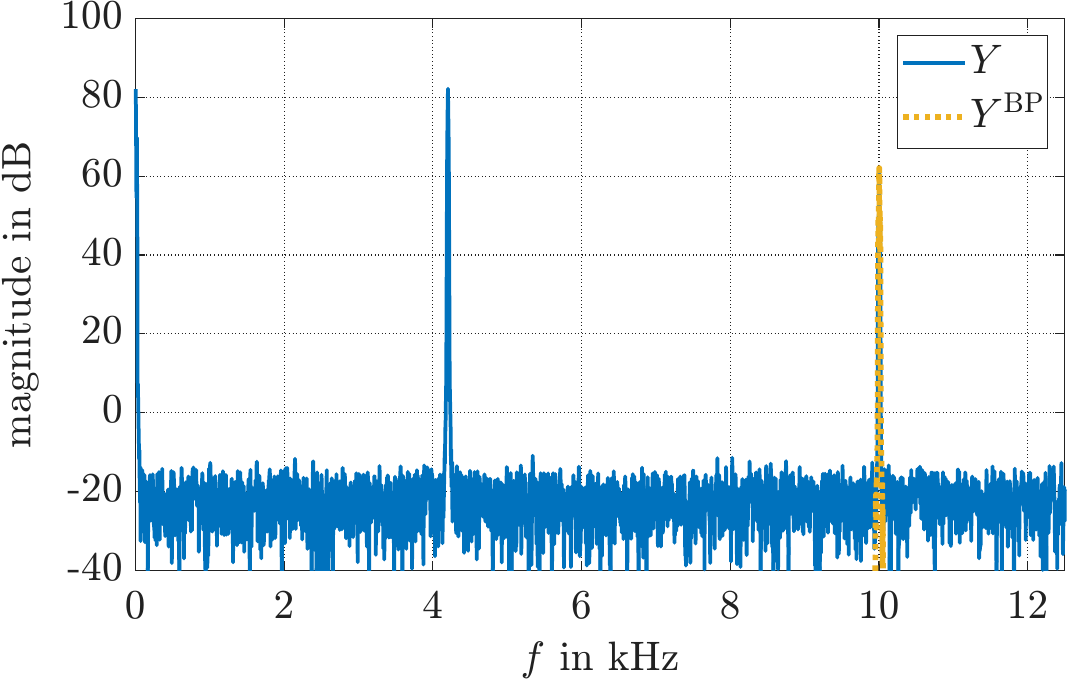}}\\
	\caption{Simulated measurement data with superposition of of three CW signals. (a)~Noisy measurement signal and extracted channel transfer function $h_3^{\mathrm{BP}}(t)$ together with its deviation from the reference. (b)~SNR of raw measurement data and of the BP-filtered transfer function. (c)~Discrete Fourier transform of the signal $y(t)$ and of its BP-filtered version.}
	\label{sim3}
\end{figure}

In a second simulation example, we superimpose three CW signals with base band frequencies $f_1 = 0$\,Hz, $f_2 = 4200$\,Hz, and $f_3=10\,000$\,Hz, all with $m_{1/2/3}(t)=1$. The resulting signal is again sampled in the base band with a time step $\Delta t = 1$\,$\mu$s corresponding to an IF or measurement bandwidth of 1\,MHz. Both $x(t)$ and $y(t)$ are superimposed with random noise and the channel transfer functions are assumed to be
\begin{eqnarray}
h_1(t)&=& 0.05\, (2 + \cos(100\, t/\mathrm{s}))\,,\notag\\
h_2(t)&=& 0.05\, (2 + \sin(50\, t/\mathrm{s}))\,,\notag\\
h_3(t)&=& 0.005\, (2 + \cos(70\, t/\mathrm{s}))\,.\notag 
\end{eqnarray}
The relevant simulation results are depicted in Fig.\,\ref{sim3}. The raw measurement signal corresponding to the measurement bandwidth of 1\,MHz is again $y(t)/x(t)$ and the BP-filtered measurement signal obtained by our signal processing methodology with a bandwidth of 120\,Hz is $h_3^{\mathrm{BP}}(t)$. Both are found in Fig.\,\ref{sim3}(a), where the observed relative deviation of $h_3^{\mathrm{BP}}(t)$ with respect to the reference $h_3(t)$ on the order of about $-60$\,dB and below corresponds again approximately to the worst-case SNR of $h_3^{\mathrm{BP}}(t)$ as found in Fig.\,\ref{sim3}(b). The worst-case SNR improvement due to the BP filtering as can be observed in Fig.\,\ref{sim3}(b) corresponds again pretty well with the expected value of  $10\log_{10}(1\,\mathrm{MHz}/120\,\mathrm{Hz}) = 39.0$\,dB. The strong peaks of the SNR in the slope regions of $h_3^{\mathrm{BP}}(t)$ are again just artifacts in the SNR estimation. In Fig.\,\ref{sim3}(c), the Fourier transforms of $y(t)$ and of its BP-filtered version are shown. The BP bandwidth is chosen that it captures the relevant signal contributions of $h_3(t)$. Remarkable is that the filtering process works still very well, even though the signal $y_3(t)$ is about 20\,dB weaker than the other two signal components. By further simulations, it was verified that the reconstruction errors tend to vanish completely if the superimposed noise is decreased, where the error contributions due to the signal processing procedure, i.e., mostly truncation errors, must also be kept small by choosing sufficiently long time sequences of the signals. However, it should also be noticed that the overall measurement signal $y(t)/x(t)$ as seen in Fig.\,\ref{sim3}(a) has strong temporal variations, which may not be desirable for all antenna measurements. As shown before, such temporal variations can be avoided by working with just one CW signal, where the measurement acceleration based on the presented signal processing procedure still works very well. 
\begin{figure}[t!]
	\centering 
	\includegraphics[width=0.84\columnwidth]{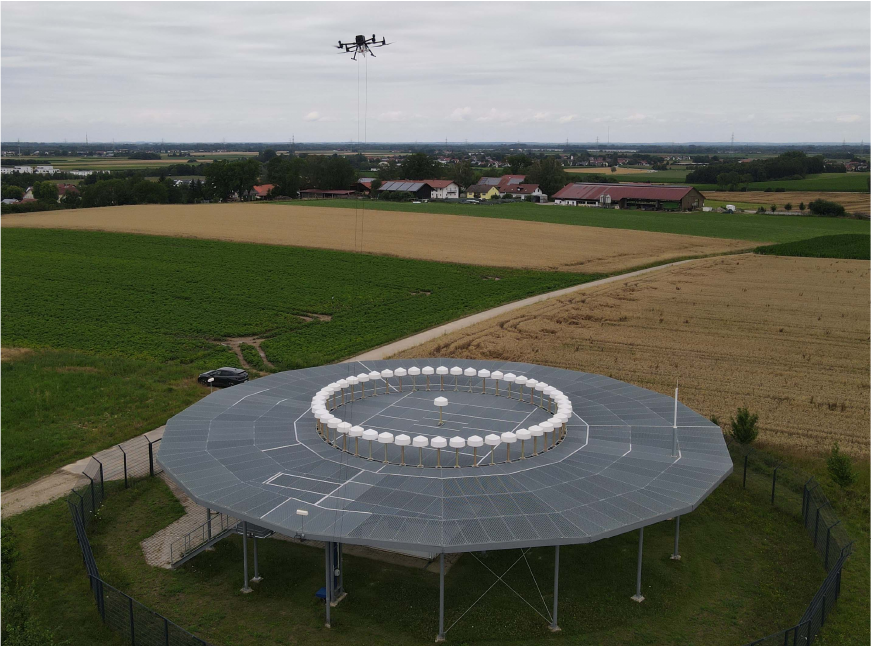}
	\caption{Photograph of the DVOR in Ottersberg, Germany, with the measurement UAV in the air.}
	\label{dvor}
\end{figure}
\begin{figure}[t!]
	\centering 
	\includegraphics[width=0.84\columnwidth]{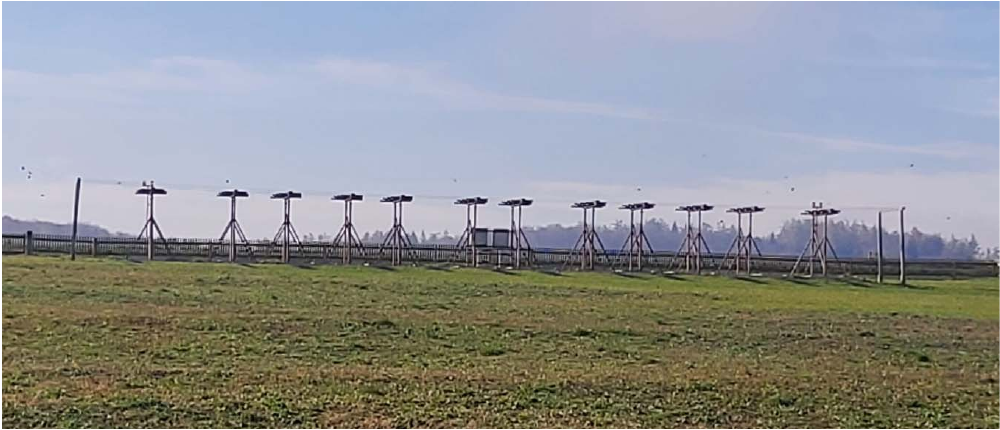}
	\caption{Photograph of the antenna array of the ILS localizer at the airport in Oberpfaffenhofen, Germany.}
	\label{ils}
\end{figure}
\begin{figure}[t!]
	\centering 
	\includegraphics[width=0.64\columnwidth]{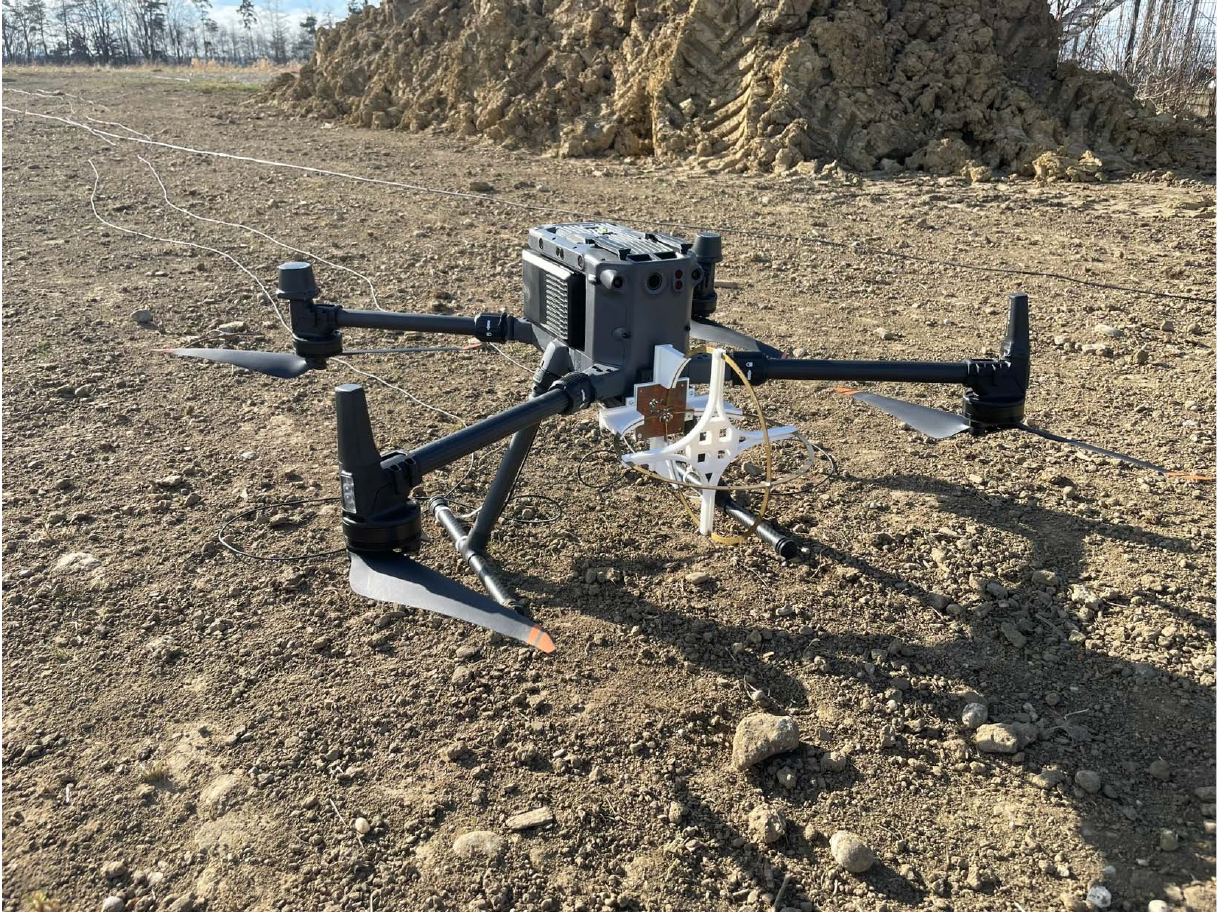}
	\caption{DJI Matrice M350 RTK UAV with installed 112\,MHz linearly dual-polarized loop antenna.}
	\label{dji}
\end{figure}
\begin{figure}[t]
	\centering
	\subfigure[~]{\includegraphics[scale=0.43,keepaspectratio]{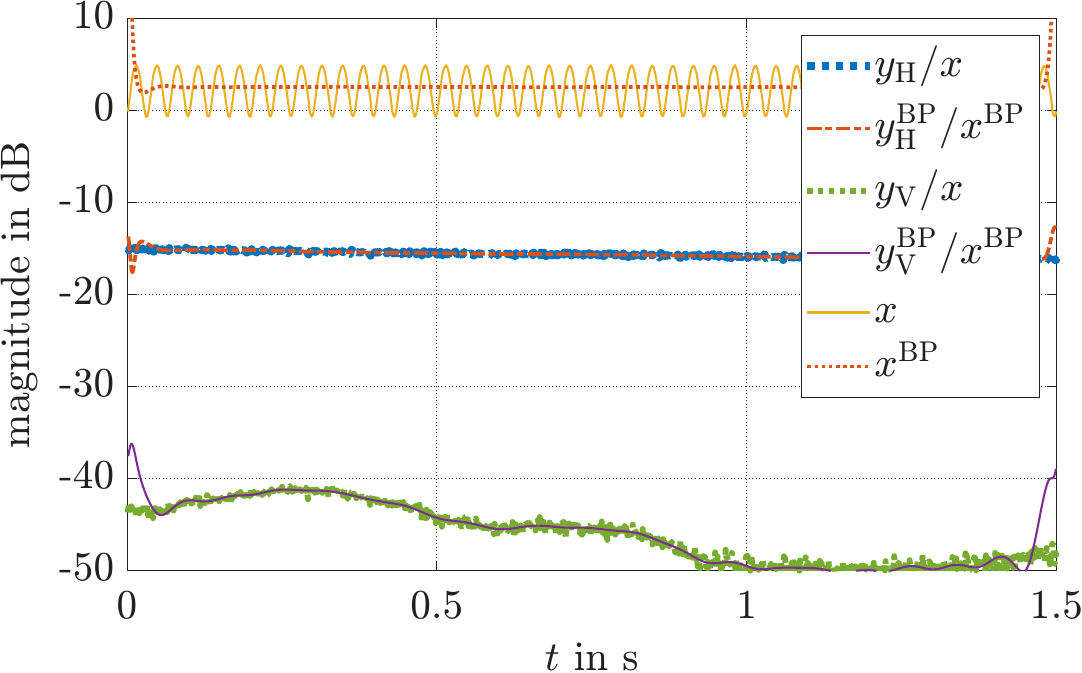}}\\[1.9mm]
	\subfigure[~]{\includegraphics[scale=0.43,keepaspectratio]{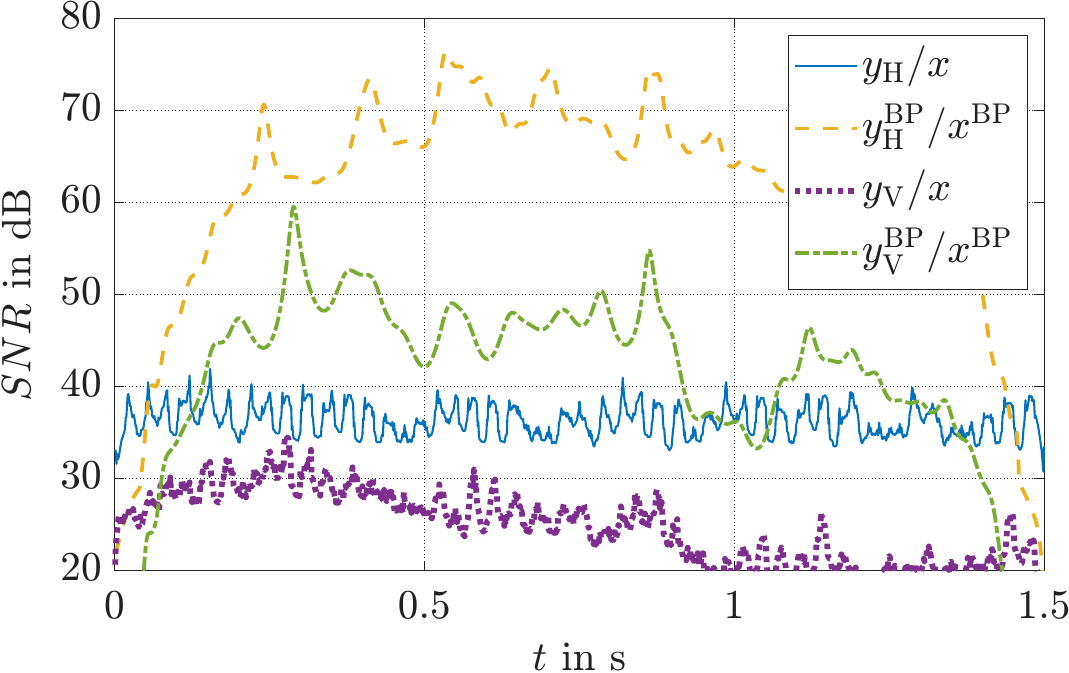}}\\[1.9mm]
	\subfigure[~]{\includegraphics[scale=0.43,keepaspectratio]{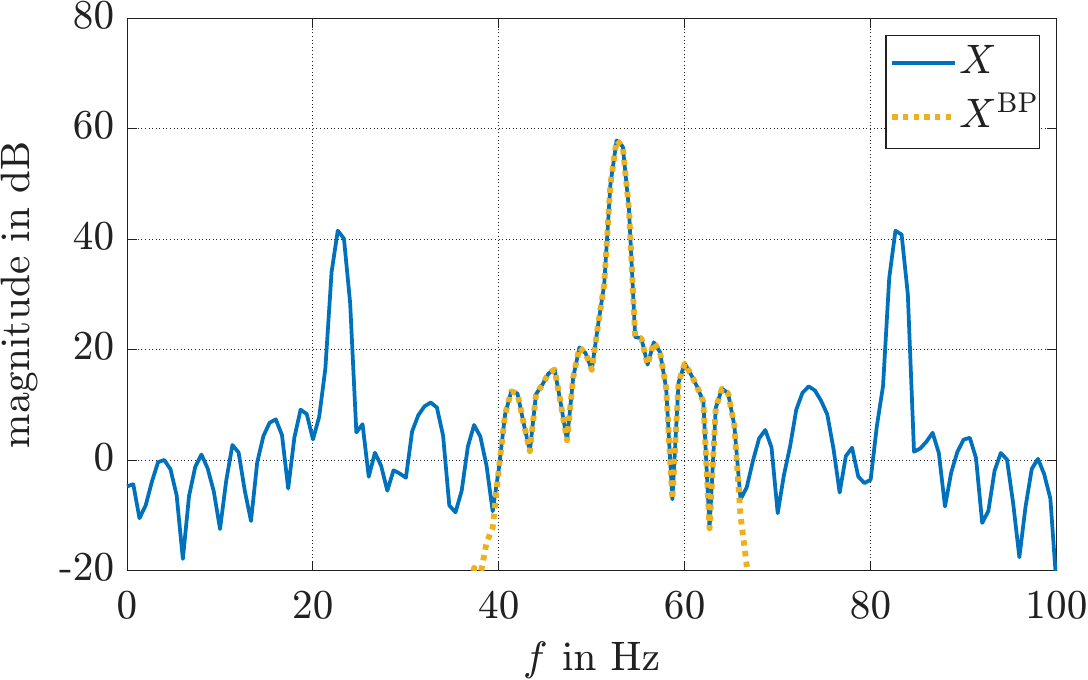}}\\
	\caption{Exemplary measurement data of the AM carrier of the DVOR. (a)~Raw measurement signals and BP-filtered measurement signals for horizontal and vertical polarization, together with the corresponding reference signals. (b)~SNR of raw measurement data and of BP-filtered measurement data. (c)~Discrete Fourier transform of the reference signal and of its BP-filtered version.}
	\label{meas1}
\end{figure}
\begin{figure}[t]
	\centering
	\subfigure[~]{\includegraphics[scale=0.43,keepaspectratio]{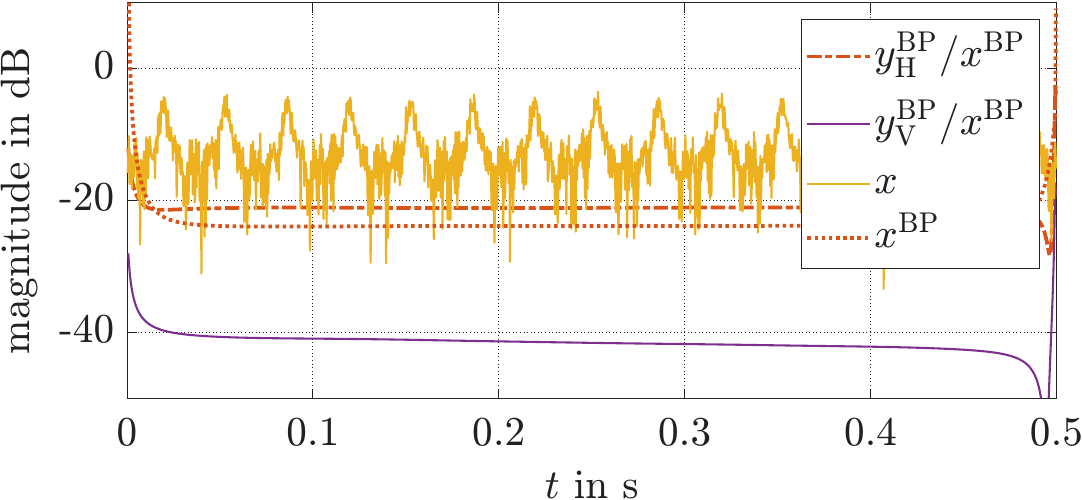}}\\[1.9mm]
	\subfigure[~]{\includegraphics[scale=0.43,keepaspectratio]{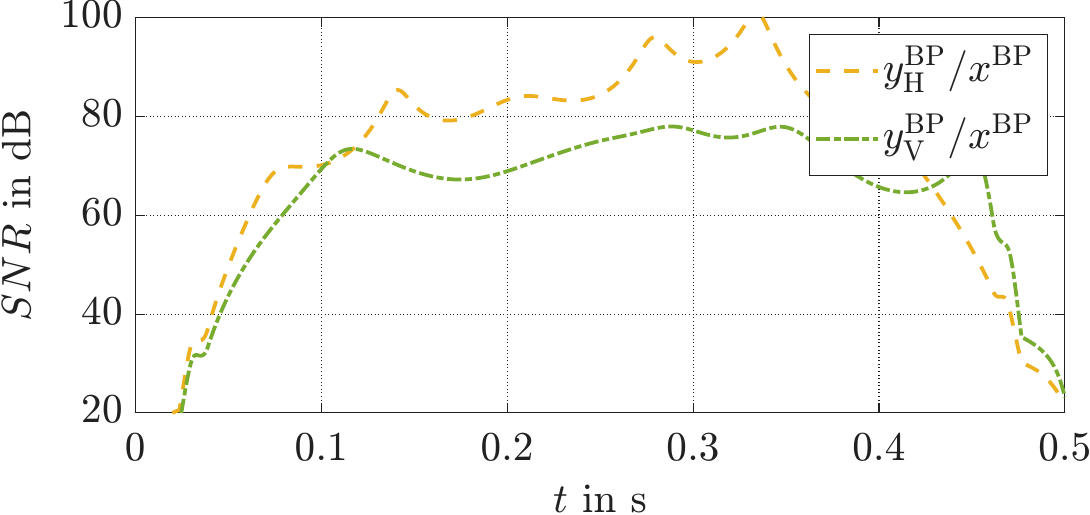}}\\[1.9mm]
	\subfigure[~]{\includegraphics[scale=0.43,keepaspectratio]{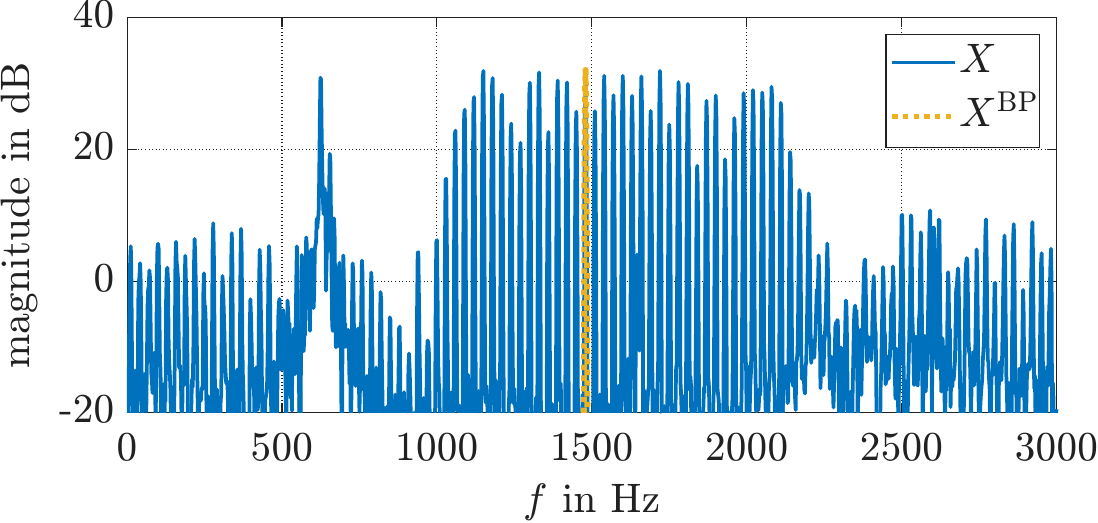}}\\
	\caption{Exemplary measurement data of the FM carrier of the DVOR. (a)~Raw measurement signals and BP-filtered measurement signals for horizontal and vertical polarization, together with the corresponding reference signals. (b)~SNR of BP-filtered measurement data. (c)~Discrete Fourier transform of the reference signal and of its BP-filtered version.}
	\label{meas2}
\end{figure}

\section{Measurement Results}
\label{results}
To further demonstrate and investigate the functionality of the presented methodologies, measurements of the time-modulated signals of terrestrial navigation systems are considered. The considered navigation systems are  the Doppler very high frequency omnidirectional radio range (DVOR) in Ottersberg, Germany, with a carrier frequencies of 112.3\,MHz as seen in Fig.\,\ref{dvor} and the localizer of the instrument landing system (ILS) at the airport in Oberpfaffenhofen, Germany, with a carrier frequency of 110.5\,MHz as seen in Fig.\,\ref{ils}. The measurements have been performed with the UAV as shown in Fig.\,\ref{dji}. The UAV carries a dual-polarized crossed-loop antenna as measurement probe, which is discussed in detail in~\cite{Salah2024,Salah2025}. The measurements have been performed with a reference antenna with a fixed location on the ground according to the configuration in Fig.\,\ref{conf}(b). More information on performing such measurements is found in~\cite{Eibert2023a}, where, however, a UAV only with a horizontally polarized probe antenna was used. Moreover, the fiber optical connections between the ground and the UAV as used in~\cite{Eibert2023a} have been replaced by coaxial cable connections for the measurements considered in this article. 
\begin{figure}[t]
	\centering
	\subfigure[~]{\includegraphics[scale=0.56,keepaspectratio]{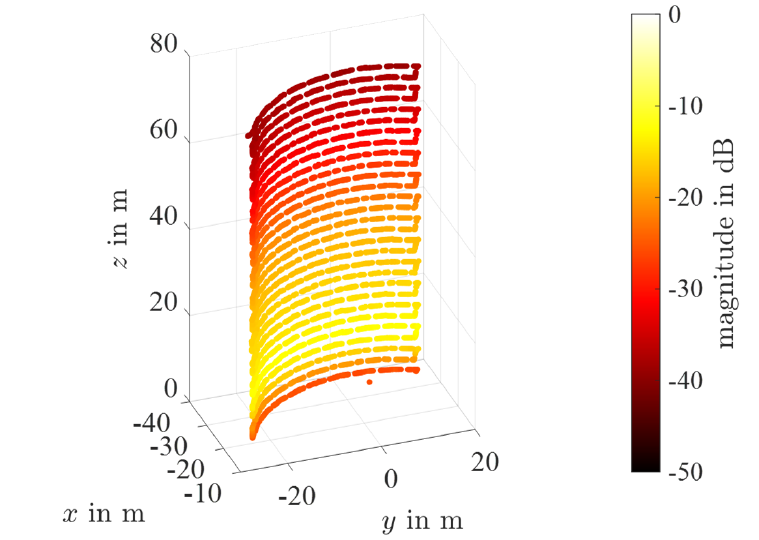}}\\[1.9mm]
	\subfigure[~]{\includegraphics[scale=0.56,keepaspectratio]{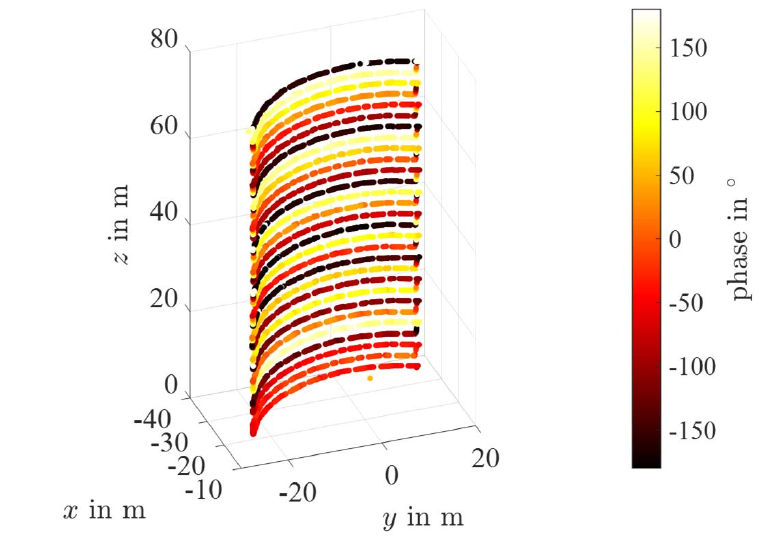}}\\
	\caption{Measured NF data of the AM carrier of the DVOR after BP filtering close to the DVOR seen in Fig.\,\ref{dvor}, which is located in the coordinate center. (a)~Magnitude of the dominant horizontal polarization. (b)~Phase of the horizontal polarization.}
	\label{dvor_am_nf}
\end{figure}
\begin{figure}[t]
	\centering
	\subfigure[~]{\includegraphics[scale=0.56,keepaspectratio]{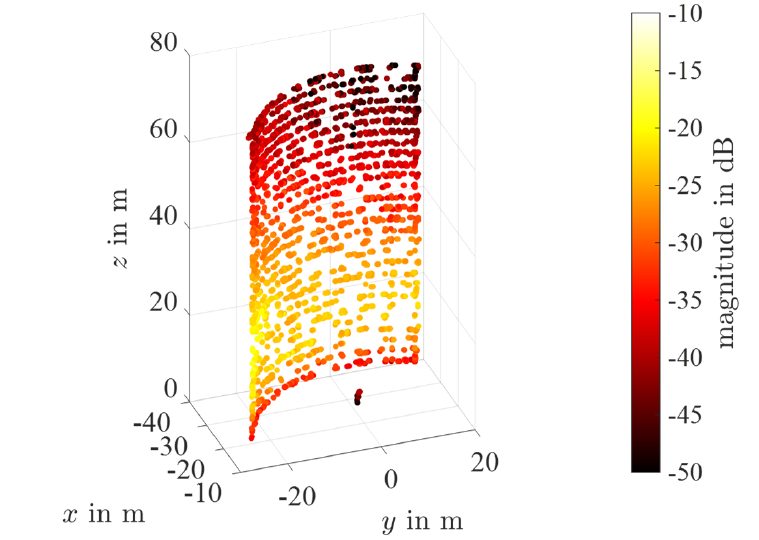}}\\[1.9mm]
	\subfigure[~]{\includegraphics[scale=0.56,keepaspectratio]{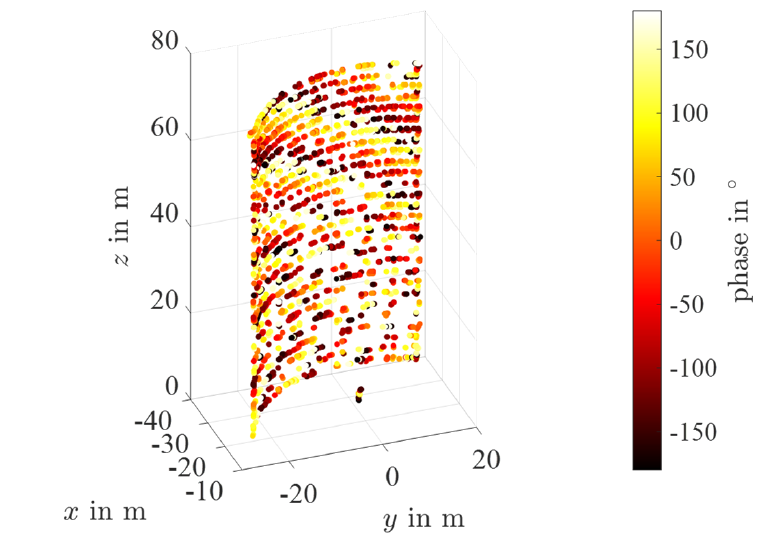}}\\
	\caption{Measured NF data of the FM carrier of the DVOR a certain modulation state close to the DVOR seen in Fig.\,\ref{dvor}, which is located in the coordinate center. (a)~Magnitude of the dominant horizontal polarization. (b)~Phase of the horizontal polarization.}
	\label{dvor_fm_nf}
\end{figure}
\begin{figure}[t]
	\centering
	\subfigure[~]{\includegraphics[scale=0.31,keepaspectratio]{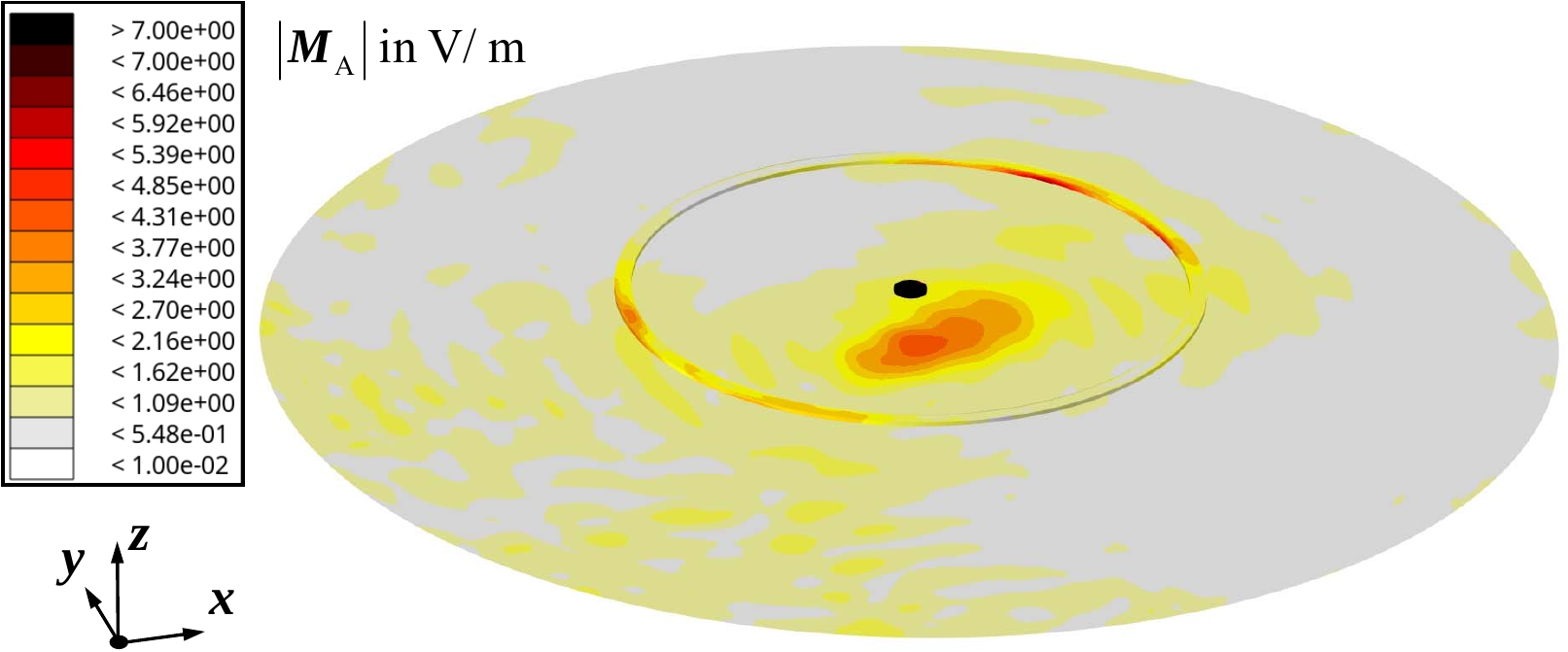}}\\[1.9mm]
	\subfigure[~]{\includegraphics[scale=0.31,keepaspectratio]{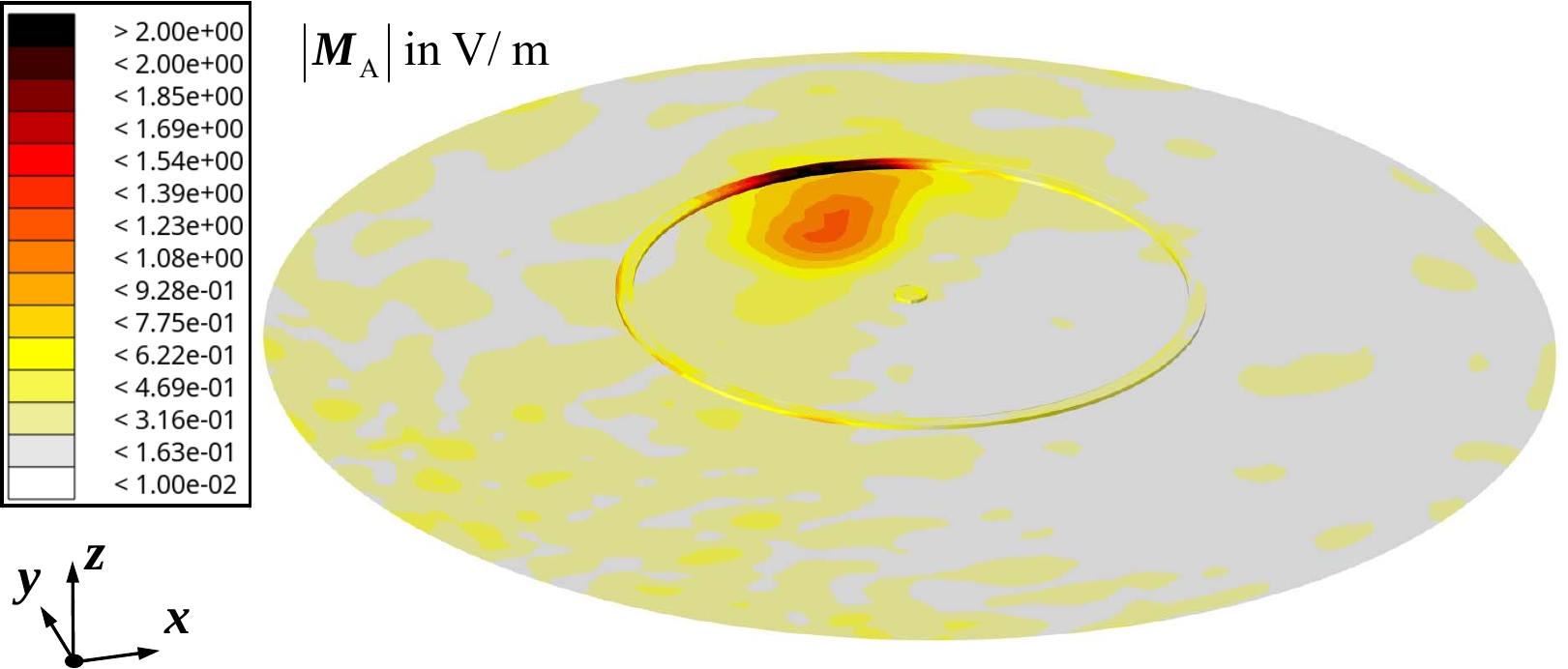}}\\
	\caption{Magnitudes of magnetic Love surface current densities on the utilized  DVOR source model obtained from inverse source solutions, (a)~for the AM carrier NF data is seen in Fig.\,\ref{dvor_am_nf}, (b)~for the FM carrier data as seen in Fig.\,\ref{dvor_fm_nf}.}
	\label{dvor_love}
\end{figure}
\begin{figure}[t]
	\centering
	\includegraphics[scale=0.5,keepaspectratio]{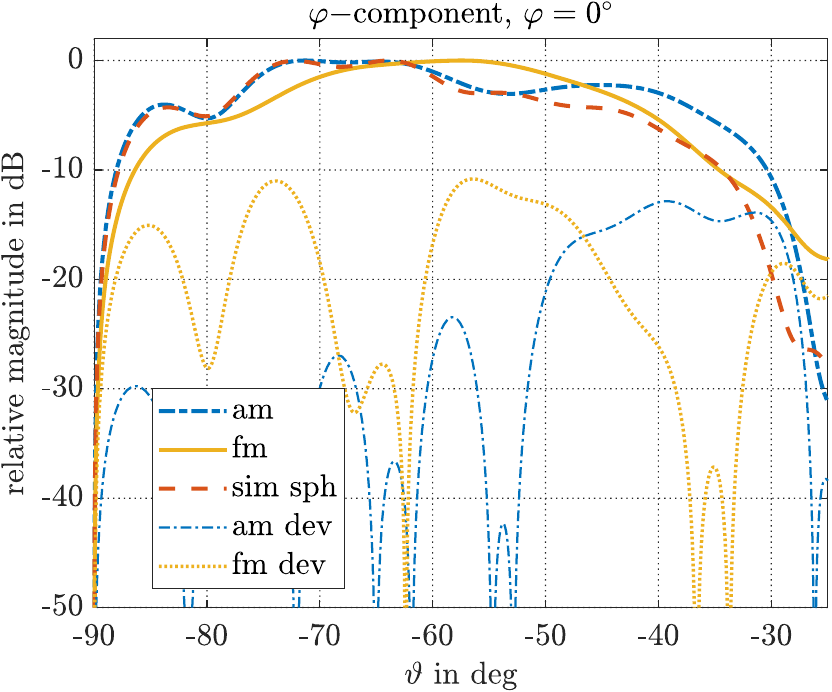}
	\caption{Vertical horizontal polarization FF pattern cut of the DVOR obtained from NFFF transformation of the measured NF data for the AM ('am') and for the FM ('fm') carrier. 'sim sph' corresponds to simulated comparison data for the center antenna element as discussed in \cite{Eibert2023a} and 'dev' indicates the deviation with respect to this comparison data. 
	}
	\label{dvor_ff}
\end{figure}
\begin{figure}[t]
	\centering
	\subfigure[~]{\includegraphics[scale=0.46,keepaspectratio]{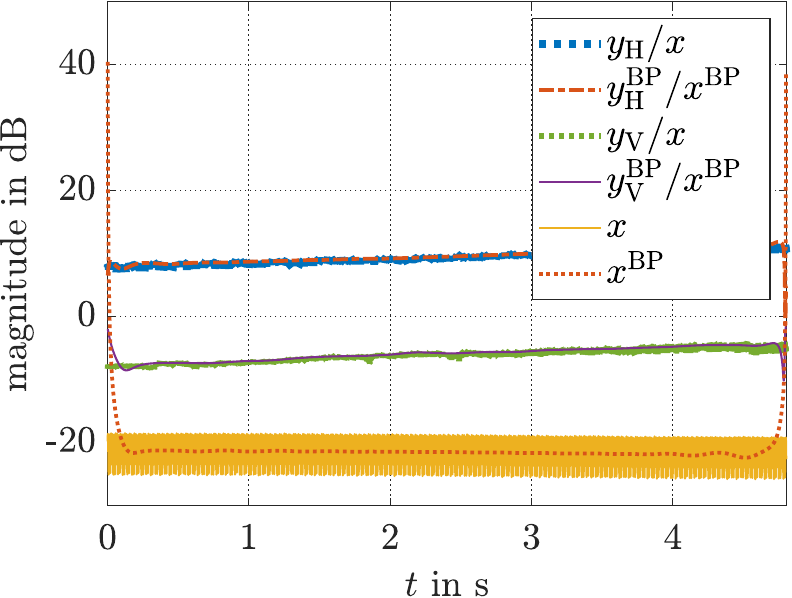}}\\[1.9mm]
	\subfigure[~]{\includegraphics[scale=0.46,keepaspectratio]{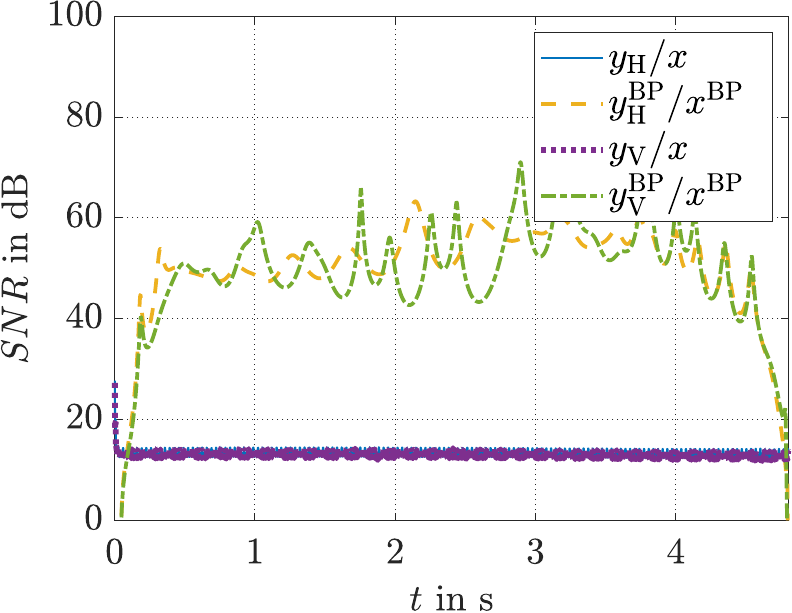}}\\[1.9mm]
	\subfigure[~]{\includegraphics[scale=0.46,keepaspectratio]{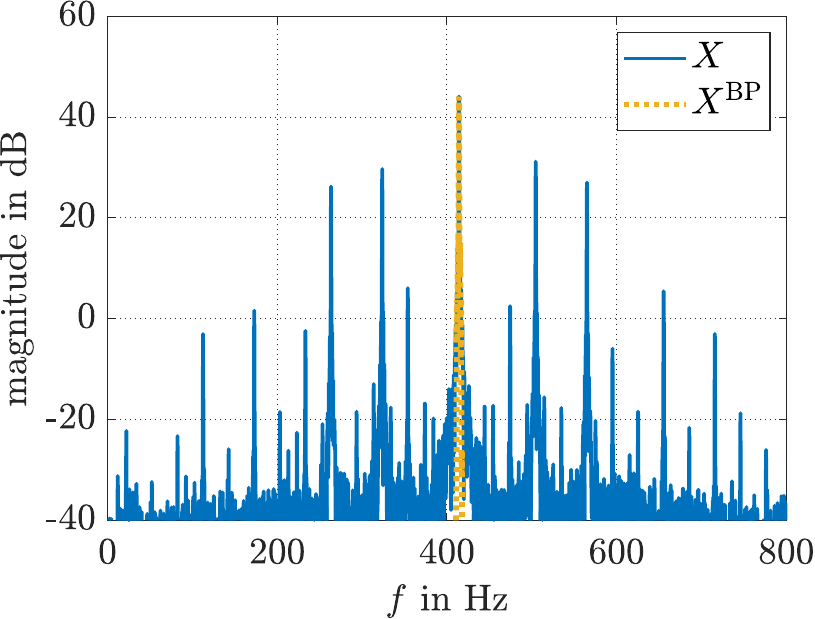}}\\
	\caption{Exemplary measurement data of the ILS localizer. (a)~Raw measurement signals and BP-filtered measurement signals for horizontal and vertical polarization, together with the corresponding reference signals. (b)~SNR of BP-filtered measurement data. (c)~Discrete Fourier transform of the reference signal and of its BP-filtered version.}
	\label{meas3}
\end{figure}
\begin{figure}[t]
	\centering
	\subfigure[~]{\includegraphics[scale=0.56,keepaspectratio]{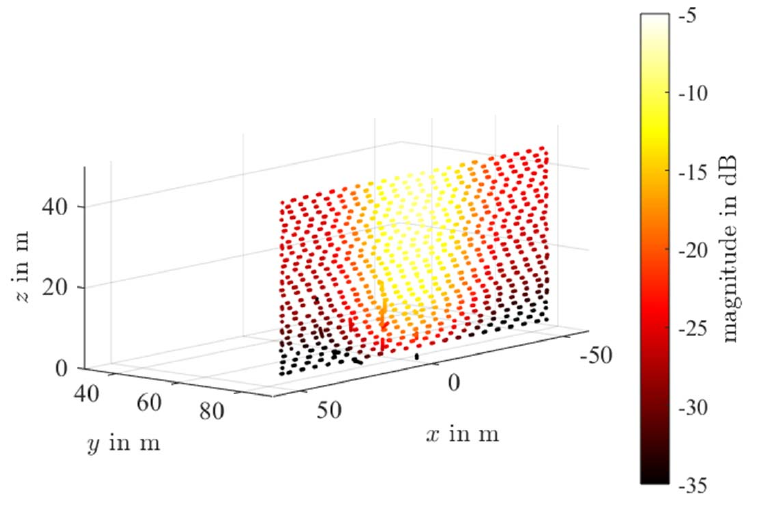}}\\
	\subfigure[~]{\includegraphics[scale=0.56,keepaspectratio]{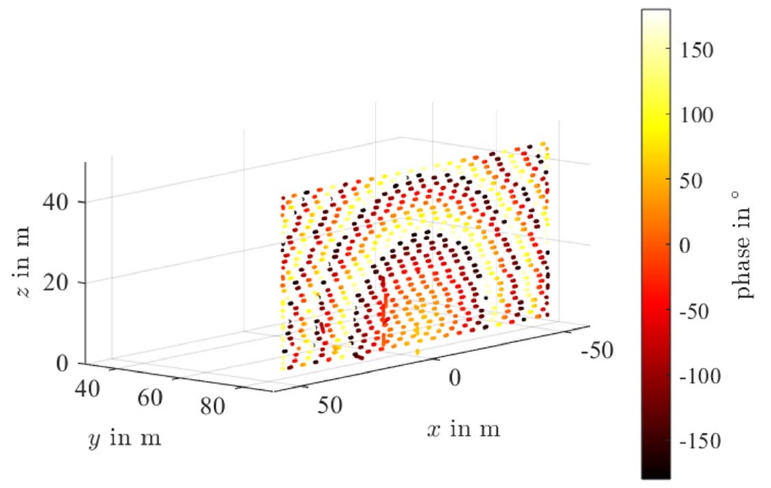}}\\
	\caption{Measured NF data of the carrier of the ILS localizer after BP filtering in about 150\,m in front of the antenna array as seen in Fig.\,\ref{ils}. (a)~Magnitude of the dominant horizontal polarization. (b)~Phase of the horizontal polarization.}
	\label{m6_c_nf}
\end{figure}
\begin{figure}[t]
	\centering
	\subfigure[~]{\includegraphics[scale=0.56,keepaspectratio]{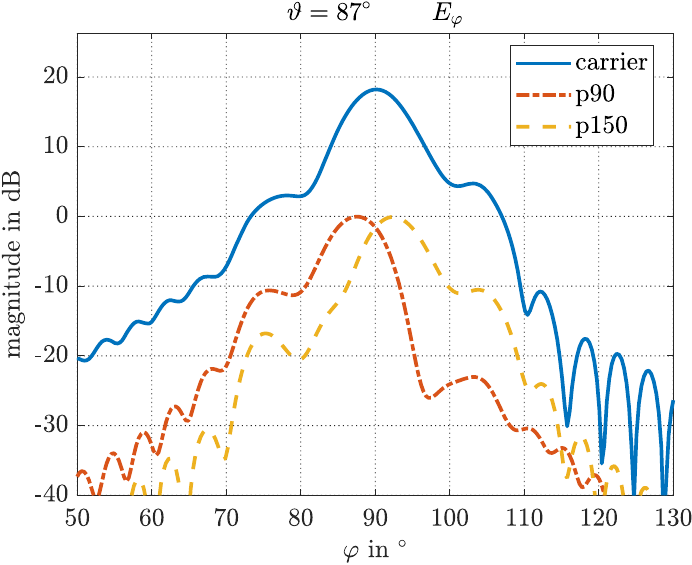}}\\[1.9mm]
	\subfigure[~]{\includegraphics[scale=0.56,keepaspectratio]{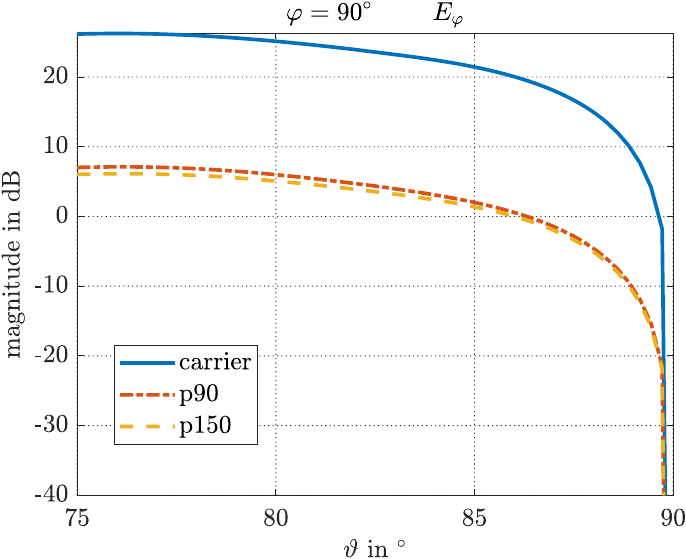}}\\
	\caption{FF pattern cuts of dominant horizontal polarization of the ILS localizer obtained from NFFF transformation of the measured NF data. (a)~Horizontal cut. (b)~Vertical cut.}
	\label{m6_ff}
\end{figure}

A first set of measurements has been collected for the 30\,Hz AM carrier transmitted by the center antenna of the DVOR. An exemplary measurement signal sequence is shown in Fig.\,\ref{meas1}.
Fig.\,\ref{meas1}(a) shows the raw measurement signals and the corresponding BP-filtered measurement signals for horizontal (subscript H) and vertical (subscript V) polarization together with the raw and BP-filtered reference signal. The related SNRs showing the improvement due to the BP filtering are depicted in Fig.\,\ref{meas1}(b). The discrete Fourier transforms of the reference signal and its BP-filtered version as given in Fig.\,\ref{meas1}(c) nicely shows the carrier and its two side bands in a distance of around 30\,Hz. The measurements have here been performed with an IF bandwidth of 1\,kHz and the bandwidth of the BP was around 30\,Hz as seen in Fig.\,\ref{meas1}(c).

A second set of measurements has been collected for the 30\,Hz FM sub-carrier transmitted by the antenna ring of the DVOR, where the FM modulation is obtained by a rotation of the radiation center around the antenna ring. At a single instance of time, just one or maybe two of the antennas around the ring are simultaneously radiating. An exemplary measurement signal sequence for these measurements is shown in Fig.\,\ref{meas2}.
Here, Fig.\,\ref{meas2}(a) shows only the BP-filtered measurement signals for horizontal (subscript H) and vertical (subscript V) polarization for a single selected spectral line together with the raw and BP-filtered reference signal, since the raw measurement signals in relation to the reference signal are not very meaningful. The related SNRs are found in Fig.\,\ref{meas2}(b). The discrete Fourier transforms of the reference signal and its BP-filtered version as given in Fig.\,\ref{meas2}(c) shows the many spectral lines due to the FM modulation with frequency separations of around 30\,Hz. The measurements have here been performed with an IF bandwidth of 3\,kHz and the bandwidth of the BP was around 30\,Hz.

The obtained NF data of the BP-filtered AM carrier as transmitted by the central antenna element of the DVOR is shown in Fig.\,\ref{dvor_am_nf}. The expected horizontally omnidirectional radiation is nicely seen in the data and also the magnitude and phase variations in vertical direction are reasonable. Corresponding NF data related to the FM carrier is found in Fig.\,\ref{dvor_fm_nf}. This data set was obtained by a BP filter with a bandwidth of about 900\,Hz around the center of the FM signal and by sorting the obtained samples according to their modulation states in the time domain as suggested for the short-time measurement approach as discussed in~\cite{Faul2023}. Shown is finally the NF for one specific temporal modulation state, where it can be expected that the radiation center is localized at a certain position on the antenna ring of the DVOR as seen in Fig.\,\ref{dvor}. The magnitude of the NF data is here not omnidirectional anymore, but the phase exhibits still a certain structure. 

For both NF data sets, inverse source solutions have been obtained based on a radiation model of the DVOR with electric $\vec{J}_{\rm A}$ and magnetic $\vec{M}_{\rm A}$ surface current densities by utilizing the inverse source solver according to \cite{Eibert2009,Eibert2015}, which can handle the irregularly sampled NF data and which is also able to consider a ground half space \cite{Parini2020,Eibert2019}. For the AM carrier, it is possible to perform CW inverse source solutions for both, the raw measurement data and the BP-filtered measurement data, since the AM modulation signal is easily removed by taking the ratio of the probe signals with respect to the reference signal. Doing so, the inverse source solution with the raw data produces an NF reconstruction error of $-24.6$\,dB, while the corresponding error obtained with the BP-filtered data is $-25.1$\,dB. Given the fact that the measurement data is influenced by a large number of error contributions, which are not that easy to control in UAV-based measurements, the improvement by $0.5$\,dB is actually not too bad. 

Love magnetic surface current densities obtained by post-processing of the inverse source solutions are depicted in Fig.\,\ref{dvor_love} for both data sets and it is very well seen that the radiation center of the AM carrier is the central antenna element of the DVOR and the radiation center of the FM signal is well localized on the antenna ring. Even better localization of the radiation centers will, of course, be obtained, if NF data on a closed cylindrical surface around the DVOR and not only a cylindrical sector as seen in Fig.\,\ref{dvor_am_nf} and Fig.\,\ref{dvor_fm_nf} would have been used. A vertical FF pattern cut obtained from the inverse source solutions of both data sets is shown in Fig.\,\ref{dvor_ff} for both linear polarizations. Both polarizations are compared to AM pattern cuts obtained from simulated hemispherical ('sim sph') data as discussed in~\cite{Eibert2023a}. For the measured AM data set, the agreement with the simulation data ('am dev') is just remarkable, whereas the agreement for the FM data set is not that good, as can certainly be expected due to the radiation center on the antenna ring.

A third set of measurements has been collected for the localizer signal of the ILS as seen in Fig.\,\ref{ils}. This signal consists of a CW carrier with AM of 90\,Hz and of 150\,Hz, where the radiation main beams of the carrier and of the AM side bands have slightly different horizontal directions. An exemplary measurement signal sequence for these measurements is shown in Fig.\,\ref{meas3}.
Fig.\,\ref{meas3}(a) shows the raw and the BP-filtered measurement signals for horizontal (subscript H) and vertical (subscript V) polarization together with the raw and BP-filtered reference signal. The related SNRs are found in Fig.\,\ref{meas3}(b). The discrete Fourier transforms of the reference signal and its BP-filtered version as given in Fig.\,\ref{meas3}(c) shows the dominant carrier, which has been filtered out by the BP, together with the symmetrically arranged side bands due to the 90\,Hz and the 150\,Hz AM. The measurements have here been performed with an IF bandwidth of 1\,kHz and the bandwidth of the BP is around 10\,Hz.

The obtained NF data of the BP-filtered AM carrier as transmitted by the ILS localizer is shown in Fig.\,\ref{m6_c_nf}. As seen, the measurement data has been collected in a vertical plane in a distance of about 150\,m in front of the antenna array as found in Fig.\,\ref{ils}. The radiation main beam is nicely seen in the data and also the phase variations are as expected. Inverse source solutions have here been obtained for the carrier NF data and also for NF data extracted for the modulation side bands, now based on a radiation model with distributed spherical harmonics sources \cite{Eibert2015} around the antenna array and again with consideration of a planar ground half space.  
The horizontal and vertical FF pattern cuts for the dominant horizontal polarization obtained from the inverse source solutions are shown in Fig.\,\ref{m6_ff}. The slightly different main beam directions of the carrier and of the modulation side bands at $+90$\,Hz ('p90') and at $+150$\,Hz ('p150') are clearly seen. The missing symmetry of the horizontal cut is due to the truncation of the NF data and due to the not perfectly planar ground.

\section{Conclusion}
\label{conclusions}
A measurement approach for performing high-SNR antenna measurements with short measurement times per measurement sample was presented and demonstrated. The key idea is to first perform the measurements with short measurement times and consequently low SNR. The 
SNR is next improved by transforming the typically equidistantly sampled measurement data into the frequency domain, perform a BP filtering of the desired signal component, and transform it back into the time domain. The noise contributions outside the bandwidth of the BP are, thus, removed, however, the relevant measurement signal variations are retained. This measurement signal processing approach can directly be employed for single-frequency measurement signals, where the bandwidth of the BP must just be large enough to keep the desired signal variations due to the changing position of the probe with respect to the AUT. However, this signal processing approach can also be utilized for multiple-frequency measurement signals and for periodically modulated measurement signals, if the bandwidth of the BP can be chosen so narrow that the desired discrete frequencies can be separated without removing the relevant measurement signal variations due to the changing position of the probe with respect to the AUT.

The approach was first demonstrated for simulated measurement data and next employed for the processing of modulated NF measurement data collected via a UAV at a DVOR and at an ILS localizer. Typical measurement signals were shown together with an increase of their SNRs by BP filtering, and single-frequency NF data sets were extracted for subsequent inverse source solutions. The diagnostic NF data and the FF results obtained from the inverse source solutions clearly demonstrated the functional principles of the DVOR and of the ILS.


%





\ifCLASSOPTIONcaptionsoff
  \newpage
\fi


\begin{thebibliography}{10}
\providecommand{\url}[1]{#1}
\csname url@samestyle\endcsname
\providecommand{\newblock}{\relax}
\providecommand{\bibinfo}[2]{#2}
\providecommand{\BIBentrySTDinterwordspacing}{\spaceskip=0pt\relax}
\providecommand{\BIBentryALTinterwordstretchfactor}{4}
\providecommand{\BIBentryALTinterwordspacing}{\spaceskip=\fontdimen2\font plus
\BIBentryALTinterwordstretchfactor\fontdimen3\font minus
  \fontdimen4\font\relax}
\providecommand{\BIBforeignlanguage}[2]{{%
\expandafter\ifx\csname l@#1\endcsname\relax
\typeout{** WARNING: IEEEtran.bst: No hyphenation pattern has been}%
\typeout{** loaded for the language `#1'. Using the pattern for}%
\typeout{** the default language instead.}%
\else
\language=\csname l@#1\endcsname
\fi
#2}}
\providecommand{\BIBdecl}{\relax}
\BIBdecl

\bibitem{Parini2020}
C.~Parini, S.~F. Gregson, J.~McCormick, D.~J.~V. Rensburg, and T.~F. Eibert,
  \emph{Theory and Practice of Modern Antenna Range Measurements, 2nd
  Edition}.\hskip 1em plus 0.5em minus 0.4em\relax London, UK: IET SciTec
  Publishing, 2020.

\bibitem{IEEE2021}
\emph{\BIBforeignlanguage{english}{{IEEE} Recommended Practice for Antenna
  Measurements}}, IEEE Std. 149-2021, Rev. R2021, 2021.

\bibitem{IEEE2012}
\emph{{IEEE} Recommended Practice for Near-Field Antenna Measurements}, IEEE
  Std. 1720-2012, Rev. R2012, 2012.

\bibitem{Yaghjian1986}
A.~D. Yaghjian, ``\BIBforeignlanguage{english}{An overview of near-field
  antenna measurements},'' \emph{\BIBforeignlanguage{english}{{IEEE} Trans.
  Antennas Propag.}}, vol.~34, no.~1, pp. 30--45, Jan. 1986.

\bibitem{Jongh1997}
R.~V.~D. Jongh, M.~Hajian, and L.~P. Ligthart,
  ``\BIBforeignlanguage{english}{Antenna time domain measurement techniques},''
  \emph{\BIBforeignlanguage{english}{{IEEE} Antennas Propag. Mag.}}, vol.~39,
  no.~10, pp. 7--11, Oct. 1997.

\bibitem{Serhir2012}
M.~Serhir, D.~Picard, F.~Jouvie, R.~Guinvarc’h, and N.~Ribiere-Tharaud,
  ``\BIBforeignlanguage{english}{Development of pulsed antennas measurement
  facility: Near field antennas measurement in time domain},'' in
  \emph{\BIBforeignlanguage{english}{Proc. 15th Int. Symp. Antenna Technol.
  Appl. Electromagn.}}, Jun. 2012.

\bibitem{Hassett2011}
K.~Hassett and B.~Williams, ``\BIBforeignlanguage{english}{Multiplexed pulsed
  transmit and receive {RF} measurement system for active phased array
  testing},'' in \emph{\BIBforeignlanguage{english}{Proc. Ann. Symp. Antenna
  Measurement Techn. Assoc.}}, Nov. 2011.

\bibitem{Faul2019}
F.~T. Faul, J.~Kornprobst, T.~Fritzel, H.-J. Steiner, R.~Strauß, A.~Wei\ss,
  R.~Geise, and T.~F. Eibert, ``\BIBforeignlanguage{english}{Near-field
  measurement of continuously modulated fields employing the time-harmonic
  near- to far-field transformation},'' \emph{\BIBforeignlanguage{english}{Adv.
  Radio Sci.}}, vol.~17, pp. 83--89, 2019.

\bibitem{Faul2023}
F.~T. Faul, J.~Daubmeier, and T.~F. Eibert,
  ``\BIBforeignlanguage{english}{Measurement and transformation of continuously
  modulated fields using a short-time measurement approach},''
  \emph{\BIBforeignlanguage{english}{Adv. Radio Sci.}}, vol.~20, pp. 9--15,
  2023.

\bibitem{Bello1963}
P.~A. Bello, ``\BIBforeignlanguage{english}{Characterization of random
  time-variant linear cbannels},'' \emph{\BIBforeignlanguage{english}{{IEEE}
  Trans. Commun. Syst.}}, vol.~11, no.~12, pp. 360--392, 1963.

\bibitem{Parsons2000}
J.~D. Parsons, \emph{\BIBforeignlanguage{english}{The Mobile Radio Propagaton
  Channel, 2nd Ed.}}, 2nd~ed.\hskip 1em plus 0.5em minus 0.4em\relax Chicester,
  UK: John Wiley \& Sons, 2000.

\bibitem{Wiener1949}
N.~Wiener, \emph{\BIBforeignlanguage{english}{Extrapolation, Interpolation, and
  Smoothing of Stationary Time Series}}.\hskip 1em plus 0.5em minus 0.4em\relax
  Cambridge, MA: The M.I.T. Press, 1949.

\bibitem{Kailath1974}
T.~Kailath, ``\BIBforeignlanguage{english}{A view of three decades of linear
  filtering theory},'' \emph{\BIBforeignlanguage{english}{{IEEE} Trans. Inf.
  Theory}}, vol.~20, no.~3, pp. 146--181, Mar. 1974.

\bibitem{Widrow1975}
B.~Widrow, J.~John R.~Glover, J.~M. McCool, J.~Kaunitz, C.~S. Williams, R.~H.
  Hearn, J.~R. Zeidler, J.~Eugene~Dong, and R.~C. Goodlin,
  ``\BIBforeignlanguage{english}{Adaptive noise cancelling: Principles and
  applications},'' \emph{\BIBforeignlanguage{english}{Proc. {IEEE}}}, vol.~21,
  no.~23, pp. 26\,848--26\,858, Dec. 1975.

\bibitem{Hu2021}
W.~Hu, Y.~Liu, W.~Si, Z.~Yao, P.~Zhao, and L.~P. Ligthart,
  ``\BIBforeignlanguage{english}{Denoising with inherent scattering
  characteristic for {2-D} radar cross section measurement},''
  \emph{\BIBforeignlanguage{english}{{IEEE} Sensors J.}}, vol.~63, no.~12, pp.
  1692--1716, Dec. 2021.

\bibitem{Harris1978}
F.~J. Harris, ``\BIBforeignlanguage{english}{On the use of windows for harmonic
  analysis with the discrete {F}ourier transform},''
  \emph{\BIBforeignlanguage{english}{Proc. {IEEE}}}, vol.~66, no.~1, pp.
  51--83, Jan. 1978.

\bibitem{Salah2024}
A.~Z. Salah and T.~F. Eibert, ``\BIBforeignlanguage{english}{Electrically small
  antennas as probe for {UAV}-based antenna measurements},'' in
  \emph{\BIBforeignlanguage{english}{Proc. German Microw. Conf.}}, Duisburg,
  Germany, 2024, pp. 65--68.

\bibitem{Salah2025}
A.~Z. Salah, D.~Unruh, and T.~F. Eibert,
  ``\BIBforeignlanguage{english}{Optimized integration of a {VHF}
  dual-polarized probe antenna for {UAV}-based measurements},'' in
  \emph{\BIBforeignlanguage{english}{Eur. Wireless}}, Sophia-Antipolis, France,
  2025, pp. 1--4.

\bibitem{Eibert2023a}
T.~F. Eibert, S.~Punzet, T.~Mittereder, F.~T. Faul, and A.~H. Paulus,
  ``{UAV}-based near-field measurements at a {D}oppler very high frequency
  omnidirectional radio range,'' in \emph{Proc. Eur. Conf. Antennas Propag.},
  Florence, Italy, 2023.

\bibitem{Eibert2009}
T.~F. Eibert and C.~H. Schmidt, ``\BIBforeignlanguage{english}{Multilevel fast
  multipole accelerated inverse equivalent current method employing
  {Rao-Wilton-Glisson} discretization of electric and magnetic surface
  currents},'' \emph{\BIBforeignlanguage{english}{{IEEE} Trans. Antennas
  Propag.}}, vol.~57, no.~4, pp. 1178--1185, Apr. 2009.

\bibitem{Eibert2015}
T.~F. Eibert, E.~Kilic, C.~Lopez, R.~A. Mauermayer, O.~Neitz, and
  G.~Schnattinger, ``\BIBforeignlanguage{english}{Electromagnetic field
  transformations for measurements and simulations (invited paper)},''
  \emph{\BIBforeignlanguage{english}{Prog. Elecromagn. Res.}}, vol. 151, pp.
  127--150, 2015.

\bibitem{Eibert2019}
T.~F. Eibert, ``\BIBforeignlanguage{english}{Efficient inverse sources
  solutions above lossy dielectric halfspace},'' in
  \emph{\BIBforeignlanguage{english}{Proc. URSI Intern. Symp. Electromagn.
  Theory}}, San Diego, CA, 2019.

\end{thebibliography}
\end{document}